\documentstyle[psfig]{mn}

\newcommand{\markcite}[1]{}

\begin{document}

\title[Cluster evolution in non-Gaussian models]
{Evolution of the cluster abundance in non-Gaussian models}

\author[J. Robinson and J.~E. Baker] {James
Robinson\thanks{jhr@palan.berkeley.edu} and Jonathan
E.~Baker\thanks{jbaker@draco.berkeley.edu}\\ Department of Astronomy,
University of California, Berkeley CA 94720}

\maketitle

\begin{abstract}
We carry out N-body simulations of several non-Gaussian structure
formation models, including Peebles' isocurvature cold dark matter
model, cosmic string models, and a model with primordial voids. We compare
the evolution of the cluster mass function in these simulations with
that predicted by a modified version of the Press-Schechter
formalism. We find that the Press-Schechter formula can accurately fit
the cluster evolution over a wide range of redshifts for all of the
models considered, with typical errors in the mass function of less
than $25\%$, considerably smaller than the amount by which predictions for
different models may differ. This work demonstrates that the Press-Schechter
formalism can be used to place strong model independent constraints on
non-Gaussianity in the universe.
\end{abstract}


\section{Introduction}
Many cosmological models, including `inflation', predict that the
large scale structure of the universe arose from the gravitational
collapse of Gaussian primordial fluctuations. These models have a
strong physical motivation (in inflation, the primordial perturbations
are just quantum fluctuations), require few free parameters (the
two-point correlation function provides a complete statistical
description of the field), and have many observable properties which
can be modeled with simple physics. Given these attractive features,
Gaussian models are a natural starting point for the investigation of
structure formation.

A number of theories however, including `defects'
(Kibble\markcite{K76} 1976, Vilenkin \& Shellard\markcite{VS94} 1994) and
certain exotic forms of inflation (Peebles\markcite{P83,P97,P98a,P98b}
1983, 1997, 1998a,b; La\markcite{L91} 1991; Amendola \&
Occhionero\markcite{AO91} 1991; Amendola \& Borgani\markcite{AB93}
1993), predict that the primordial fluctuations were not Gaussian. These
theories possess many of the attractive features of Gaussian models:
defect models, for instance, are well motivated and require few free
parameters. Unlike the Gaussian case, however, making robust
predictions in non-Gaussian models is often very difficult. In defect
theories, this is because the physical processes which must be
modeled are complex and highly non-linear. More generally, the
parameter space available to non-Gaussian models is infinitely larger,
because specifying the statistics of the primordial fluctuations
requires not just the two-point correlation function, but all higher
$n$-point correlation functions as well.

Previous studies of non-Gaussian structure formation (e.g., Weinberg
\& Cole\markcite{WC} 1992; Borgani et~al.\markcite{BCMP94} 1994; Park,
Spergel \& Turok\markcite{PST} 1991) have tended to concentrate on
specific models, making predictions for observable properties in each
case. Due to the infinite range of possible parameters, and the
difficulties of making robust predictions even within well specified
models, we turn here to a different approach: We start from a property
of non-Gaussian models which can be fully quantified by a small number
of parameters, the probability distribution function (PDF) of the
density field over the range of scales relevant to galaxy cluster
formation.  While knowledge of the PDF does not fully specify the
non-Gaussian nature of the distribution (which would require knowledge
of all the higher $n$-point correlation functions), we show that it
provides us with sufficient information to make robust predictions for
a very interesting observable quantity: the redshift evolution of the
cluster number abundance.  We do this by demonstrating that the
cluster evolution in non-Gaussian N-body simulations can be accurately
described by a version of the Press-Schechter formalism (Press \&
Schechter\markcite{PS} 1974) modified to allow for non-Gaussianity
(Chiu, Ostriker \& Strauss\markcite{COS} 1997), which requires only
the PDF as input. The result of this work is a powerful tool with
which we can use observations of galaxy clusters to constrain the
primordial PDF in the universe, without reference to any specific
non-Gaussian models or the uncertainties which may be inherent to
them. The work presented here validates a number of previous studies
which have used the non-Gaussian Press-Schechter formalism, but have
not tested it (Chiu et~al.\markcite{COS} 1998, Robinson, Gawiser \&
Silk\markcite{RGS} 1998, van de Bruck\markcite{vdB} 1998). Other work
(Robinson, Gawiser
\& Silk\markcite{Retal} in preparation) discusses in detail the strong
model-independent constraints we can place using this tool.

In \S\ref{sec-PS} we describe the PS formalism, and we extend the
derivation in the non-Gaussian case to allow for a scale dependent
probability distribution function. Section~\ref{sec-nongauss}
describes our non-Gaussian models and the realization of the initial
density fields in each case. In \S\ref{n-body} we give details of our
N-body simulations. Finally, in \S\ref{sec-results} we present our
results and in \S\ref{sec-conclusions} we discuss our conclusions.

\section{Modified Press-Schechter Formalism}
\label{sec-PS}
The Press-Schechter (PS) formalism\markcite{PS} (Press \& Schechter 1974)
is a simple semi-analytic tool for predicting the expected number
density of clusters in an evolved density field, given the linear
initial conditions. Extensive work has shown that this formalism can
give a good fit to the cluster abundance observed in N-body
simulations of Gaussian structure formation models.  Though the PS
formula has mainly been applied to Gaussian fluctuations, there is a
straightforward way to generalize it to the non-Gaussian case: Here,
the statistics of the underlying density field are quantified by
specifying its PDF $p_R(\delta)$, where $p_R(\delta)\,{\rm d}\delta$ is
the probability that the primordial field at a given point in space
has an overdensity between $\delta$ and $\delta+{\rm d}\delta$ after a
top hat smoothing of scale $R$. We can rewrite $p_R(\delta)$ in terms
of a rescaled PDF $P_R(y)$ which has zero mean and {\it rms\/} one:
$p_R(\delta)= P_R(\delta/\sigma_R)/\sigma_R$, where $\sigma_R$ is the
{\it rms\/} overdensity in spheres of radius $R$. For many
non-Gaussian models it should be a reasonable assumption that only the
{\it rms\/} value of $\delta$ varies as a function of scale (for
instance, in defect theories where the fluctuations are layed down in
a scale invariant manner), so that $P_R(y)$ is in fact independent of
$R$. The Press-Schechter formula for this case has been derived by
Chiu et~al. (1998) and used to make predictions for a number of
non-Gaussian scenarios (Robinson et~al. 1998\markcite{RGS}; van de
Bruck\markcite{vdB} 1998).

In this work, we rederive the formula for the more general case that
the form of the PDF does depend on scale, since we will see later
that the assumption of scale independence is valid for some but not
all of the models we consider.  The Press-Schechter formalism starts
from the assumption that clusters with mass greater than some value
$M$ at redshift $z$ form in regions of the primordial density
field whose overdensity smoothed on scale $R_{M}$ and extrapolated
using linear gravity to redshift $z$ is greater than some
critical value $\delta_{\rm c}$. The scale $R_{M}$ is chosen to be the
radius of a sphere containing mass $M$ in the primordial homogeneous
universe, which satisfies
\begin{equation}
\label{eq-mass-r}
M= \frac {4\pi}{3} \rho_{\rm m} R_{M} ^3,
\end{equation}
where $\rho_{\rm m}$ is the comoving matter density of the universe. The
critical threshold $\delta_{\rm c}$ is taken from an analytic solution for
the collapse of a spherically symmetric overdensity, which gives
$\delta_{\rm c}=1.69$ for a critical universe (for a derivation of the weak
cosmological dependence, see Lacey \& Cole\markcite{LC} 1993, Eke et
al.\markcite{ECF} 1996).

From this starting point, it is possible to derive the cluster
mass function. The probability ${\cal P}_{>R}$ of a point in space
forming part of a cluster with Lagrangian radius larger than $R$ is
equal to the probability of the density field, after smoothing on
scale $R$, having an overdensity larger than the critical
overdensity $\delta_{\rm c}$. That is,
\begin{equation}
{\cal P}_{>R}=\int_{\delta_{\rm c}}^{\infty} p_R(\delta) \,{\rm d}\delta.
\end{equation}
To obtain the probability ${\cal P}_R\,{\rm d}R$ of a point in space
forming part of a cluster with Lagrangian radius between $R$ and
$R+{\rm d}R$, we differentiate the above expression with respect to
$R$ and take the absolute value,
\begin{equation}
{\cal P}_R \,{\rm d}R = \left| \frac {\rm d} {{\rm d}R}{\cal P}_{>R}
\right| \,{\rm d}R.
\end{equation}
We can obtain the number density of such clusters $n(R)\,{\rm d}R$ by dividing
by the cluster Lagrangian volume,
\begin{equation}
\label{eq-nr}
n(R) \,{\rm d}R = \frac {3 f} {4\pi R^3} \left| \frac {\rm d} {{\rm d}R}
{\cal P}_{>R} \right| {\rm d}R,
\end{equation}
where we have also multiplied by a correction factor $f$, whose value we will
fix by ensuring that final mass function accounts for the entire mass
of the universe.
The integrated mass function $N_{>M}$ is then given by
\begin{equation}
N_{>M}= \frac {3 f} {4\pi} \int_{R_{M}}^{\infty} \frac{{\rm d}R}{R^3}
\left| \frac {\rm d} {{\rm d}R}{\cal P}_{>R} \right|.
\end{equation}
To fix the value of the correction factor $f$, we use the following
expression for the
integrated mass density $\rho_{\rm m}$ of the universe:
\begin{equation}
\rho_{\rm m}
= \int_{0}^{\infty} M_{R}\, n(R)\,{\rm d}R, 
\end{equation}  
where $M_{R}$ is the cluster mass corresponding to Lagrangian
radius $R$. Substituting using equation~\ref{eq-mass-r} and~\ref{eq-nr}, we find
\begin{equation}
\label{eq-correction}
{f}= \left(  \left. {\cal P}_{>R} \right|_{R=0}\right)^{-1}
\end{equation}
which does indeed give $f=2$ for Gaussian fluctuations, as
originally proposed by Press \& Schechter.

We comment briefly on some computational issues: We can rewrite the
mass function in a convenient form using integration by parts,
\begin{equation}
\label{eq-ngtm}
N_{>M}= \frac {3 f} {4\pi} \left| \frac{1}{R_{M}^3} {\cal
 P}_{>R_{M}} - 3 \int_{R_{M}}^{\infty} \frac{{\rm d}R}{R^4} {\cal
 P}_{>R} \right|.
\end{equation}
For each model and for each redshift we compute and store the function
${\cal P}_{>R}$ for a range of values of $R$. Each computation of
$N_{>M}$ (eq.~\ref{eq-ngtm}) then requires only a single integration
over these stored values. Finally, we note that the redshift
dependence of the mass function arises through the
dependence of ${\cal P}_{>R}$ on redshift, which for conciseness we
have not labelled explicitly here.

\section{Non-Gaussian models}
\label{sec-nongauss}
For the purposes of this work, we consider three classes of
non-Gaussian model; Peebles' isocurvature cold dark matter (ICDM)
model, cosmic string models, and primordial void models.
These models are all physically motivated, and each one represents an
interesting structure formation scenario in its own right. However,
each model has a number of uncertain features. In the cases of ICDM or
Voids, there is a wide range of possible model parameters to consider,
and in the case of cosmic strings, there are extreme computational
difficulties involved in realistically evolving a string network. Here
we will not be concerned with these uncertainties, since our goal is
simply to pick a diverse set of concrete non-Gaussian models, and use
these examples to test the applicability of the non-Gaussian PS
formalism. To the extent to which the formalism works, we will have a
tool with which we can make specific predictions for cluster evolution
in a wide class of non-Gaussian models, taking as input only the
initial PDF $p_R(\delta)$ (which is what enters into the PS
calculation; see \S\ref{sec-PS}). We will then be able to use
cluster evolution to constrain the initial PDF of the universe without
referring to any specific non-Gaussian models, a possibility which
will be extremely useful given the difficulty of making robust
predictions in so many non-Gaussian scenarios.

We now explain the motivation for each of the models, and present the
computational details of realizing initial density fields in each
case.

\subsection{Peebles ICDM model}
The first non-Gaussian structure formation scenario we consider is the
Peebles ICDM model (Peebles\markcite{P83,P97,P98a,P98b}
1983, 1997, 1998a,b). This is an inflationary model which gives rise
to non-Gaussian isocurvature fluctuations in the matter
distribution. The cold dark matter (CDM) in this model is a field
$\phi({\bf x})$ with mass density
\begin{equation} 
\rho({\bf x}) \propto \phi({\bf x}) ^2
\end{equation}
which survives as a relic at the end of inflation. Perturbations in
this CDM field grow from quantum fluctuations which are frozen during
the accelerated expansion of the universe. The perturbations in
$\phi({\bf x})$ are Gaussian, with power spectrum
\begin{equation}
P_\phi\propto k^{m_\phi}
\end{equation}
over a broad range of scales, where the spectral index $m_\phi$
depends on the details of the model.
The corresponding fluctuations in $\rho$ are non-Gaussian, and it is
easy to show that they have an initial power spectrum
\begin{equation}
P_\rho\propto k ^ {m_\rho},
\end{equation} 
with 
\begin{equation}
\label{eq-mphi}
m_\rho=3 + 2 m_\phi.
\end{equation}

Due to the isocurvature nature of the fluctuations, the CDM density
does not evolve until the matter era, and then only on scales which
are inside the horizon. Those modes which have crossed the horizon
before matter-radiation equality will all start to grow at that time,
while those modes which enter the horizon later will be increasingly
suppressed. For our investigation of cluster formation we are
interested mainly in the former, smaller scales, so it will be sufficient
to assume scale-independent linear growth of the initial
fluctuations.

We consider two values of the initial spectral index of the Gaussian
field $m_\phi$. First, we consider a model (which we denote ICDM-2.4)
with $m_\phi=-2.4$, in a background cosmology with $\Omega_{\rm m}=0.2$ and
$\Omega_\Lambda=0.8$. This set of parameters has been suggested as an
interesting starting point by Peebles (1998b) since it gives 
power spectra in reasonable agreement with large scale structure and
CMB data.  Second, we consider a model (ICDM-2.0) with $m_\phi=-2.0$,
in a background cosmology with $\Omega_{\rm m}=1$ and $\Omega_\Lambda=0$. We
do not expect this model to give a particularly good fit to the large
scale structure data, but we introduce it since the degree of
non-Gaussianity is intermediate between that of (ICDM-2.4) and a
Gaussian distribution, and it therefore allows us to test the
Press-Schechter formalism over a wider range of models.

For each ICDM model, we generate a $128^3$ realization of a Gaussian random
field $\phi({\bf x})$ with the appropriate power spectrum $P(k)\propto m_\phi$. 
We
then compute $\rho(x)=\phi(x)^2$, and the corresponding density
perturbation 
\begin{equation}
\delta(x)=\frac{\rho(x)-\bar{\rho}} {\bar{\rho}},
\end{equation}
where $\bar{\rho}$ is the mean density, given by
\begin{equation}
\bar{\rho}=\frac{1}{N} \sum_{N}\rho(x),
\end{equation}
and $N$ is the number of points on the lattice. Finally, we rescale
the density contrast so that the density field will have the correct
value of $\sigma_8$ when linearly evolved to $z=0$.

\subsection{Cosmic Strings}
The second class of non-Gaussian model we consider is that of cosmic
strings. Here, structure is seeded gravitationally by an evolving
network of string-like `defects', topological relics of a symmetry
breaking phase transition in the very early universe
(Kibble\markcite{K76} 1976; Vilenkin \& Shellard\markcite{SV94}
1994). Recent work suggests that cosmic strings in a critical universe
do not give a viable theory of structure formation (Allen et
al\markcite{Aetal1997} 1997; Albrecht, Battye \&
Robinson\markcite{ABR} 1997), although strings in a universe with a
significant cosmological constant may be able to give a good fit to
existing large scale structure data (Battye, Robinson \&
Albrecht\markcite{BRA} 1998; Avelino et~al\markcite{ASWA} 1998). For
the purposes of this work, we shall not be concerned by the many
uncertainties in the details of realistic string dynamics. Instead, we
shall take a simple representative model and use it as a testing
ground for our study of cluster evolution.

Our string-seeded density models use the string simulations of
Ferreira (1995\markcite{F95}). These simulations evolve a network of
strings in flat space, where an exact solution of the equations of
motion exists (Smith \& Vilenkin\markcite{SmV} 1987). Although the
string simulations are carried out in a flat space background, the
network evolution demonstrates many of the important properties which
are seen in expanding universe simulations, the most important of
which is the `scaling' of the string density, ensuring that the
induced density perturbations will be roughly scale invariant on the
largest scales.

Following Veeraraghavan \& Stebbins (1990\markcite{VandS}), the matter
overdensity $\delta$ induced by the network can be computed by
convolving a component $\theta_+=\theta_{00}+\theta_{ii}$ of the
string stress-energy with a suitable Green function, and integrating
over time. That is,
\begin{equation}
\label{eq-VS}
\tilde{\delta}({\bf k}) = 4\pi \int_{\eta_{\rm i}}^{\eta_{\rm f}} {\rm d}\eta' T({\bf
k},\eta') \tilde{\theta}_+({\bf k},\eta') \frac {1}{1+(k_{\rm
c}/k)^2},
\end{equation}
where $\eta_{\rm i}$ and $\eta_{\rm f}$ are the initial and final
times in the simulation, $\tilde{\delta}({\bf k})$ is the Fourier
transform of the induced overdensity, $\tilde{T}_2^C$ is a transfer
function taken from Veeraraghavan \& Stebbins (1990\markcite{VandS}),
$\eta=\int_0^t {\rm d}t /a$ is the conformal time, $a$ is the
cosmological scale factor, and the last term is a `compensation
factor', with compensation scale $k_{\rm c}=2\pi/\eta$, included to
ensure artificially that the density evolution conserves stress energy
on the largest scales.

We carry out our flat space string simulations on a $128^3$ lattice,
choosing the side of our box to be $100h^{-1}$ Mpc. We work out the
conformal time $\eta$ at each time-step by equating it to the
simulation horizon size (which is zero at the initial time and grows
by one lattice spacing at each time-step). At each step, we compute
the appropriate component $\theta_+({\bf x})$ of the string stress
energy tensor on the lattice, Fourier transform to obtain
$\tilde{\theta}_+({\bf k})$, and multiply by $T({\bf k},\eta$). The
final density contrast is then given by the sum of individual
contributions from each time-step, as in eq.~\ref{eq-VS}. Since all
significant perturbations are induced at a time when the fluctuations
are still linear, it is sufficient to use linear gravity in this way
to compute the initial density field, and only consider the effects of
non-linear gravity in the subsequent evolution. We choose two
representative string models. First, we consider a model with an
$\Omega_{\rm m}=1$ hot dark matter (HDM) background with one species of
massive neutrino accounting for the entire mass of the universe, and a
Hubble constant $H_0=100h$ km s$^{-1}$ Mpc$^{-1}$ with $h=0.5$. We
denote this model `Strings-HDM'. Second we consider a string network
in an $\Omega_{\rm m}=1$ CDM background, also with $h=0.5$, which we denote
`Strings-CDM'.

\subsection{Primordial Voids}
The final non-Gaussian model we consider is one where the matter
distribution has an enhanced network of primordial voids. The physical
motivation for such models comes from an inflationary theory with two
scalar fields (La\markcite{L91} 1991): one field drives inflation,
while at the same time the second field undergoes a first order phase
transition. The bubbles nucleated in this phase transition give rise
to under-densities in the CDM distribution, superimposed on Gaussian
fluctuations which are also generated during the inflationary
epoch. The detailed properties of the bubble distribution, including
their sizes, shapes and profiles, will depend on the exact nature of
the inflationary model, but for some range of parameters it is
possible to produce bubbles which are large enough to have a
significant impact on structure formation, yet still escape limits
arising from current Cosmic Microwave Background (CMB) observations
(Amendola \& Occhionero\markcite{AO91} 1991; Amendola \&
Borgani\markcite{AB93} 1993).

A typical cosmologically interesting scenario would produce a dense
network of bubbles with radii of order $10$--$40 h^{-1}$ Mpc. For the
purposes of this work, we generate a simple realization of such a
model as follows: in a $100h^{-1}$ Mpc box, we generate a $128^3$ realization of 
a
Gaussian density field $\delta_{\rm g} ({\bf x})$, with a power spectrum
$P_\rho\propto k^{-1.8}$. This spectral index is chosen so that the
density field has a very similar power spectrum to that in the
ICDM-2.4 model, thus ensuring that our simulations primarily compare
the effect of varying the amount of non-Gaussianity in the fluctuations,
rather than that of varying the spectral index. We then superimpose on
this a distribution of $N_{\rm V}$ randomly located voids, where the voids are
spherically symmetric top-hat under-densities with radius
$R=12h^{-1}$ Mpc. The total density field is given by
\begin{equation}
\delta({\bf x}) = \delta_{\rm g}({\bf x}) - \delta_{\rm V} \sum_{i=1}^{N_{\rm V}} 
W_R({\bf x} - {\bf y_i}),  
\end{equation} 
where $\delta_{\rm V}$ is the amplitude of the void under-density, ${\bf
y_i}$ is the centre of the $i^{\rm th}$ void and
$W_R({\bf x})$ is the window function in real space of a spherical
top-hat with radius $R$, that is:
\begin{equation}
W_R({\bf x})=\left\{ 
\begin{array}{ll}
1&\ldots|{\bf x}| \le R\\
0&\ldots\mathrm{otherwise.}
\end{array} \right.
\end{equation}
We dial the amplitude of the void under-density $\delta_{\rm V}$ to
ensure that the resulting matter field has a significant degree of
non-Gaussianity, without completely swamping the Gaussian component of
the distribution. We find $\delta_{\rm V} = 0.1 \sigma_{{\rm g},8}$ is
a suitable value, where $ \sigma_{{\rm g},8}$ is the {\it rms\/} value
of the Gaussian component, after top-hat smoothing on a scale of
$8h^{-1}$ Mpc. We choose $N_{\rm V}=12$, and generate the void centres
${\bf y_i}$ by picking points at random, discarding any choices which
would give rise to overlapping voids, and repeating until the number
of acceptable centres equals $N_{\rm V}$. Finally we rescale the field
$\delta({\bf x})$ to ensure that it has the chosen normalization
$\sigma_8$.

\section{Details of N-body runs}
\label{n-body}
In \S\ref{sec-nongauss} we have explained the motivation for our
non-Gaussian models, and set out in detail the processes involved in
generating density fields which realize these models. We now describe
the techniques involved in transforming these linear density fields
into initial conditions for N-body simulations. For each of the models
in question we start from a density field $\delta({\bf x})$ defined on
a $128^3$ lattice. The N-body simulations require initial conditions
in the form of positions and velocities of some number $N_{\rm
p}=n_{\rm p}^3$ particles, which we can generate from our density
fields using the Zeldovich approximation (Zeldovich\markcite{Z70}
1970). Specifically, we define a displacement field ${\bf p}({\bf q})$
via
\begin{equation}
\label{eq-disp}
\nabla{\bf \cdot} {\bf p} ({\bf q}) = - \delta_0 ({\bf q}),
\end{equation}
where the subscript zero denotes the density field extrapolated to the
present day assuming linear growth. Starting from a regular grid of
particles, the position ${\bf x}(z)$ of each particle at redshift $z$
is given by
\begin{equation}
{\bf x}(z)= {\bf q} + {\bf p}({\bf q}) \frac{b(z)}{b(0)}, 
\end{equation}
where $b(z)$ is the linear growth factor at redshift $z$, and $q$ is
the initial coordinate of the particle.  The comoving velocity of that
particle is given by
\begin{equation}
{\bf v}({\bf q},z) = {\bf p} ({\bf q}) \frac{b(z)}{b(0)} \frac{{\rm
d}\ln b}{{\rm d}t}
\end{equation}
where $t$ is the physical time. We can solve eq.~\ref{eq-disp} for the
displacement field efficiently using Fourier transforms, that is: 
\begin{equation}
\tilde{\bf p}({\bf k}) = {\bf G}({\bf k}) W({\bf k}) \tilde{\delta}({\bf k}), 
\end{equation}
where $\tilde{\bf p}({\bf k})$ is the Fourier transform of the
displacement field ${\bf p}({\bf x})$, $G$ is the appropriate Green function,
\begin{equation}
{\bf G}({\bf k}) = - i \frac {\bf k} {k^2},
\end{equation}
and $W({\bf k})=w(k_1)w(k_2)w(k_3)$ is a smoothing function, whose
form is chosen to optimize agreement between the power spectrum of
the input density field $\delta({\bf x})$ and that of the mass
distribution represented by the displaced particles (which we can
measure by resampling the particle distribution as a new density field
on a lattice).  We find the choice
\begin{equation} 
w(k)=
\left\{
\begin{array}{ll}
1&\ldots k L /n_{\rm p} < \pi\\
0&\ldots \mathrm{otherwise}
\end{array}
\right.
\end{equation}
gives good agreement.

We evolve each density field using P$^3$M simulations carried out
on a GRAPE-3AF board attached to a Sun workstation.  The GRAPE-3AF
is a special purpose computer which quickly evaluates pairwise forces,
allowing a cosmological simulation with $64^3$ particles to be run in
under a day.  The code used for our simulations is based on the P3MG3A 
algorithm described by Brieu, Summers, \& Ostriker\markcite{BSO95} (1995).

Each simulation contains $N_{\rm p}=n_{\rm p}^3=64^3$ particles with a particle mass 
of $1.05 \times 10^{12}\Omega_{\rm m} h^{-1}$ M$_\odot$.  The effective force 
resolution $\eta$ is 0.128 times the mean inter-particle separation, or
$200h^{-1}$ kpc for our $100h^{-1}$ Mpc box length.  The initial redshifts 
$z_{\rm i}$ for each model, chosen in each case by ensuring that $\Delta^2<0.15$ at
the Nyquist frequency, are given in Table~\ref{tab-all}. 
\begin{table*}
\begin{minipage}{110mm}
\centering
\begin{tabular}{cccccccccc}
\hline 
 &&&&&\multicolumn {2}{c}{$P(k)$}& \multicolumn {3}{c}{$P_R(y)$}\\
Model&$\Omega_{\rm m}$& $\Omega_\Lambda$ & $z_{\rm i}$ & $\sigma_8$ & $C_1$ &
$C_2$ & $A_1$ &   $A_2$ &   $A_3$\\
G$\Lambda$-1.8 & 0.2 & 0.8 & 19 & 0.98 & -1.82 & 0     & 0    & 0 & 0 \\ 
ICDM-2.4       & 0.2 & 0.8 & 39 & 0.99 & -1.60 & 0     & 0.58 & 0 & 0 \\
ICDM-2.0       & 1.0 & 0.0 & 29 & 1.0  & -0.98 & 0     & 0.35 & 0 & 0 \\
Strings-HDM    & 1.0 & 0.0 & 19 & 0.58 & -6.95 & -3.67 & 0.17 & 0 & 0 \\
Strings-CDM    & 1.0 & 0.0 & 29 & 0.86 & -1.9  & -0.6  & 0.19 & -0.012 & 0 \\
Voids          & 1.0 & 0.0 & 15 & 0.89 & -2.07 & 0     & 0.154 &-0.049 & 0.001\\
\hline
\end{tabular}
\vspace{10pt}
\caption{Summary of each of the models, including background cosmology,
normalization, power spectrum and PDF parameters.}
\label{tab-all}
\end{minipage}
\end{table*}
For a given
amplitude of fluctuations, the criterion of no shell crossing
generally forces us to adopt an earlier initial redshift in the
non-Gaussian models than in the corresponding Gaussian case. Outputs
from the simulations are stored and analyzed for a number of redshifts
between $z=z_{\rm i}$ and $z=0$.

For each output, we identify clusters in the simulations using
the `friends-of-friends' (FOF) algorithm. This algorithm defines
clusters as those groups of particles which are connected by pairs
whose separation is smaller than some linking parameter $l$, which is
quoted in units of the mean inter-particle separation. We choose
$l=0.2$, so that the mean overdensity within the clusters
identified is of order 200. This choice
of linking length has been shown to pick out clusters whose mass
function is in good agreement with the PS prediction in the case of
Gaussian fluctuations (Mo, Jing \& White\markcite{MJW} 1996).

\section{Results}
\label{sec-results}
We now compare the predictions of the modified PS formalism with the
results of our simulations. In order to compute the PS prediction, we
require as input for each model the amplitude of the fluctuations
(quantified in terms of $\sigma_8$), the form of the power spectrum
$P(k)$, and the PDF $P_R(y)$ (renormalized to have mean zero and
standard deviation one) as a function of scale. For our non-Gaussian
models we do not have simple analytic expressions for these
quantities, so instead we must measure them directly in the
simulations. To do this, we output the particle positions and
velocities at an early redshift $z_{\rm l}=(z_{\rm i}-1)/2$, when all
scales of interest are still in the linear regime. We compute the PDF
of the field on a range of scales $R$ by throwing down 30,000 spheres
at random and counting the distribution of particle numbers in each
sphere. We compute the power spectrum by resampling the particle
distribution as a density field on a $128^3$ lattice, and performing
the appropriate Fourier transform.

With Gaussian simulations, it is not normally necessary to carry out
these steps, since, in constructing a realization of a Gaussian field
it is straightforward to ensure that it has the correct normalization,
power spectrum and PDF\@. In order to test the fairness of the
techniques we are using to measure these quantities in the
non-Gaussian case, we carry out an identical analysis on a simulation
(which we denote $G\Lambda$-1.8) with Gaussian initial conditions and
a power spectrum $P(k)\propto k^{-1.8}$, in a background cosmology with
$\Omega_{\rm m}=0.2$ and $\Omega_\Lambda=0.8$. 

\begin{figure}
\centerline{\psfig{file=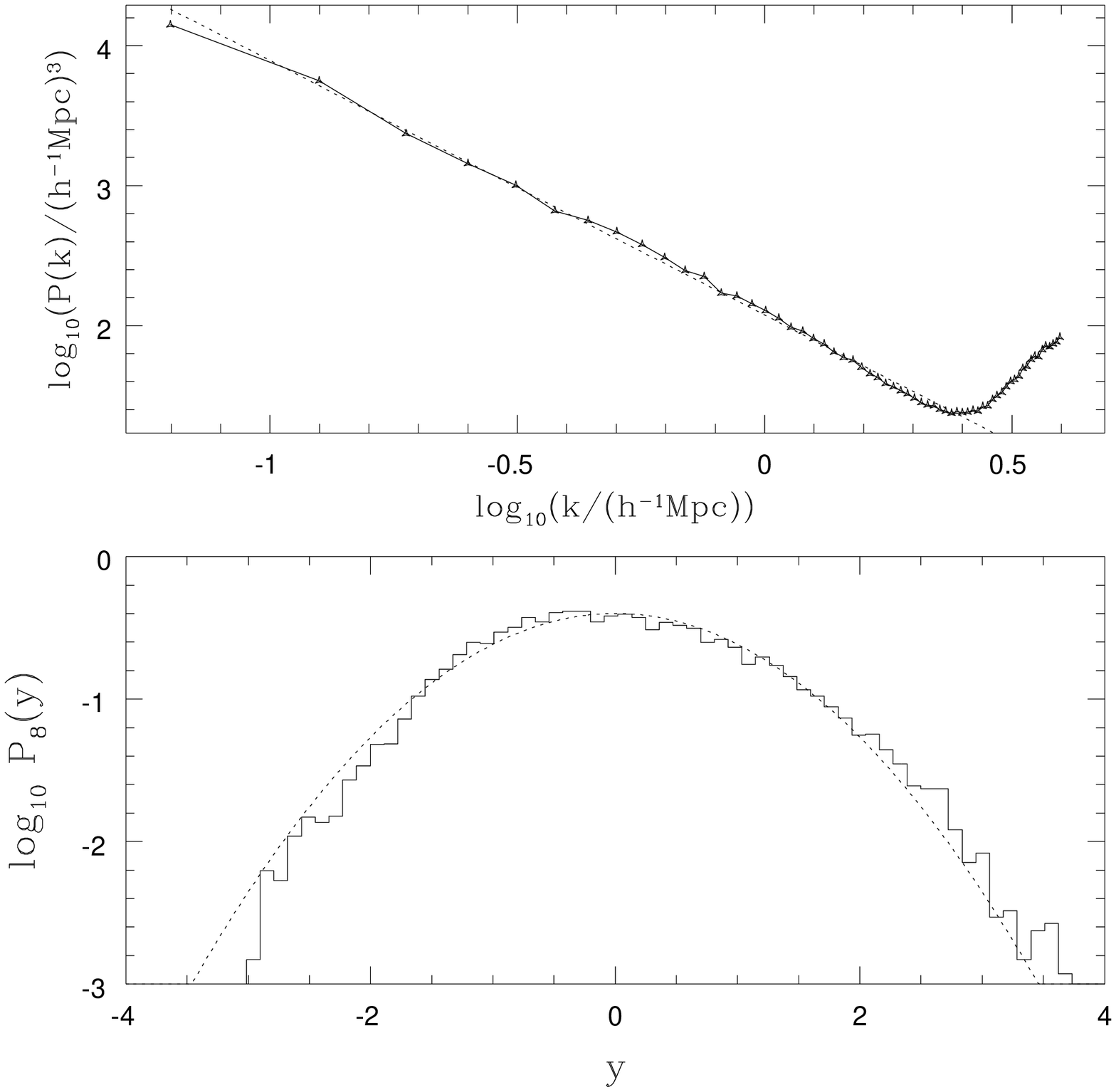,width=3.5in}}
\caption{Measured (solid lines) and fitted (dotted lines) power spectra (top
panel) and PDF (bottom panel) in the Gaussian G$\Lambda$-1.8 model.}
\label{fig-fits-gl-1.8}
\end{figure}

\begin{figure}
\centerline{\psfig{file=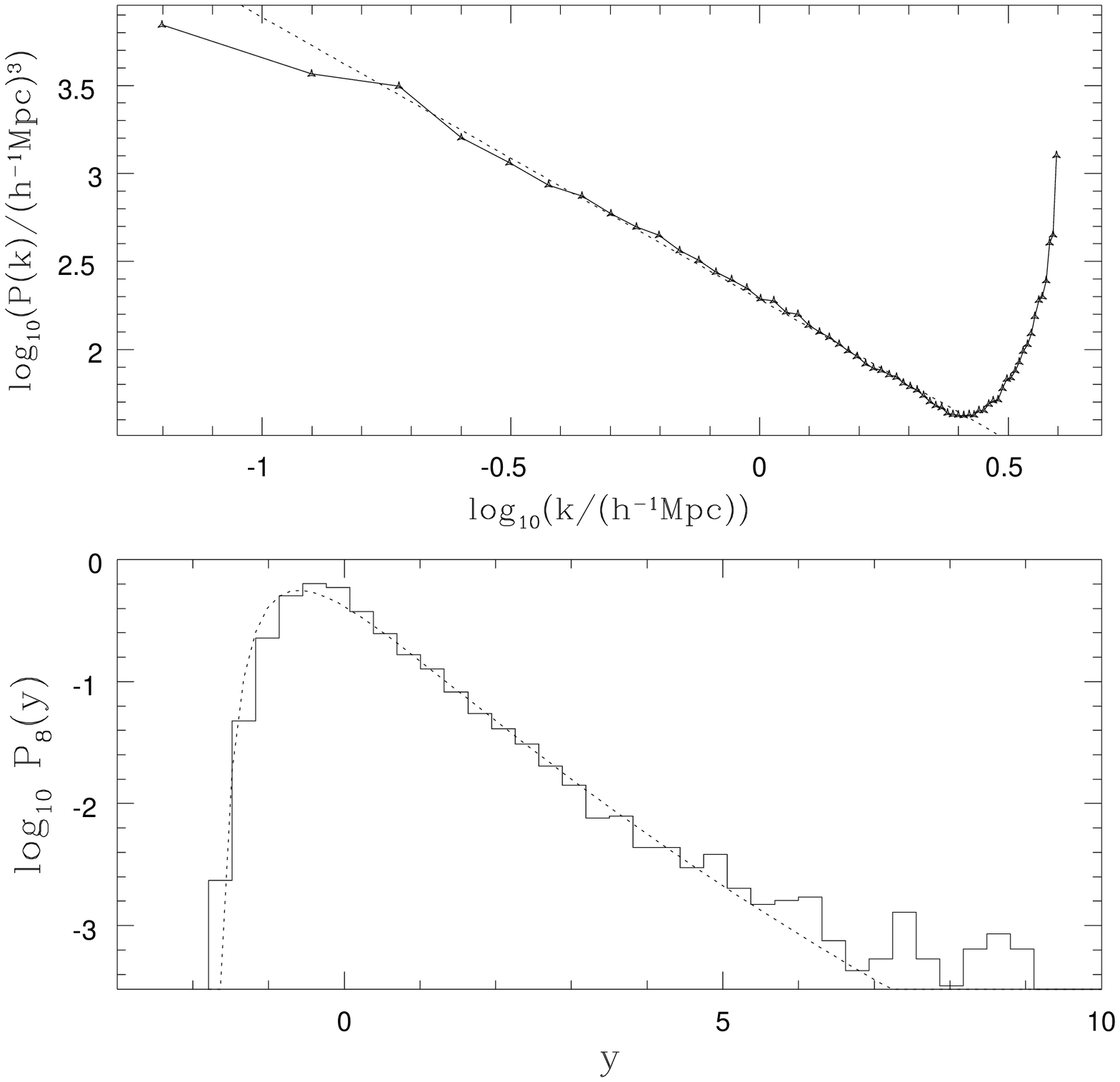,width=3.5in}}
\caption{Measured (solid lines) and fitted (dotted lines) power spectra (top
panel) and PDF (bottom panel) in the ICDM-2.4 model.}
\label{fig-fits-icdm-2.4}
\end{figure}

\begin{figure}
\centerline{\psfig{file=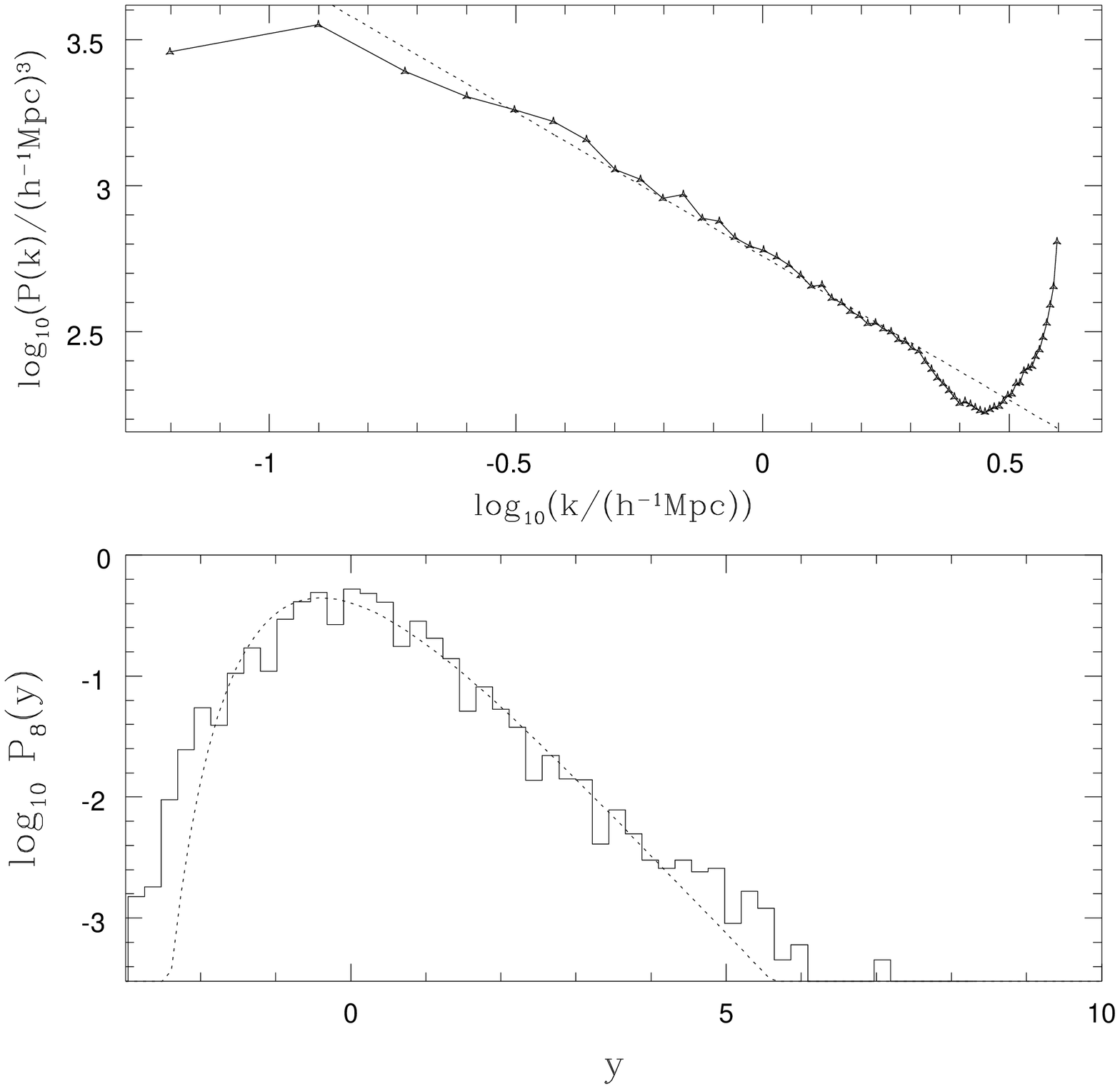,width=3.5in}}
\caption{Measured (solid lines) and fitted (dotted lines) power spectra (top
panel) and PDF (bottom panel) in the ICDM-2.0 model.}
\label{fig-fits-icdm-2.0}
\end{figure}

\begin{figure}
\centerline{\psfig{file=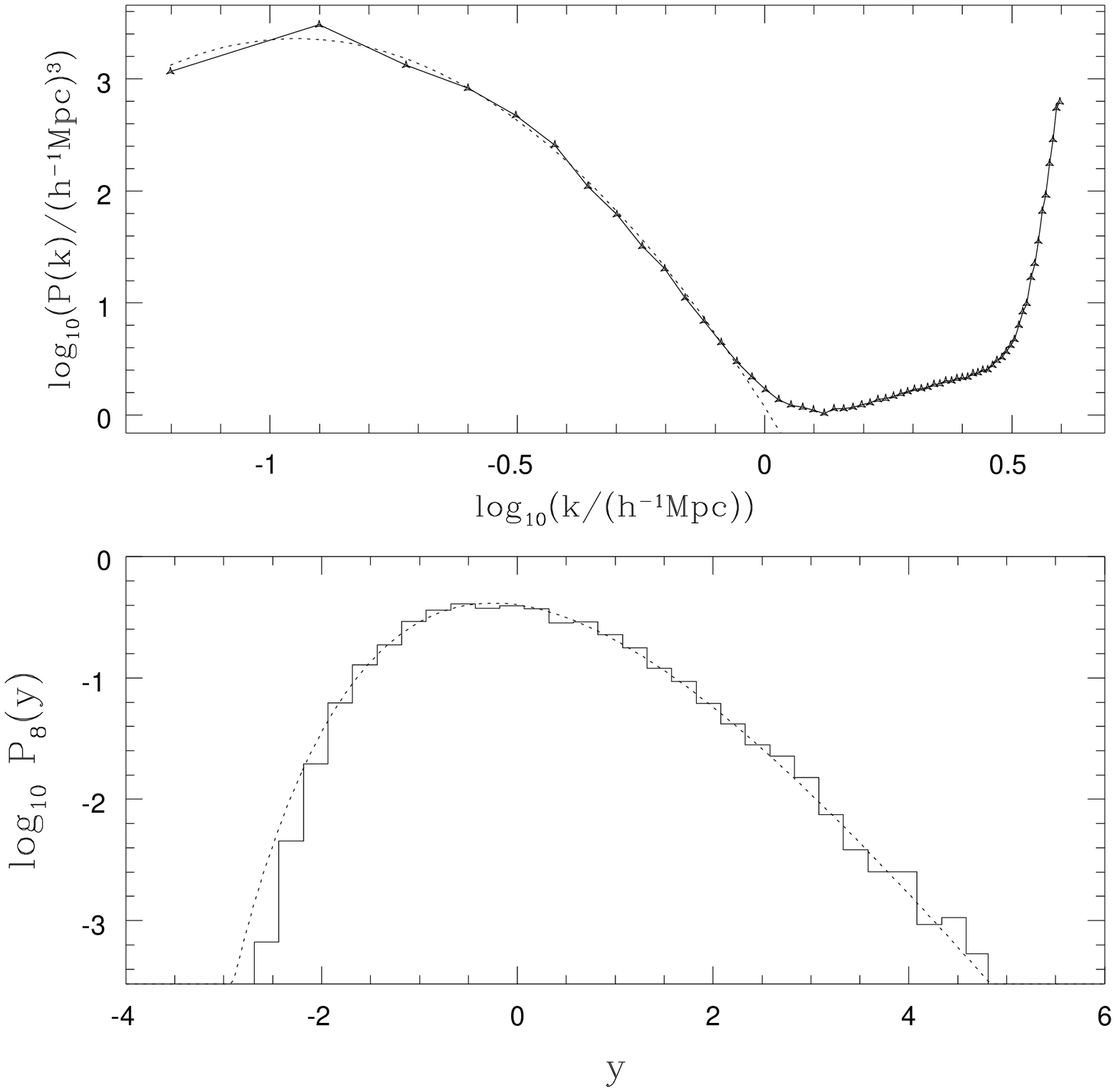,width=3.5in}}
\caption{Measured (solid lines) and fitted (dotted lines) power spectra (top
panel) and PDF (bottom panel) in the Strings-HDM model.}
\label{fig-fits-string-HDM}
\end{figure}

\begin{figure}
\centerline{\psfig{file=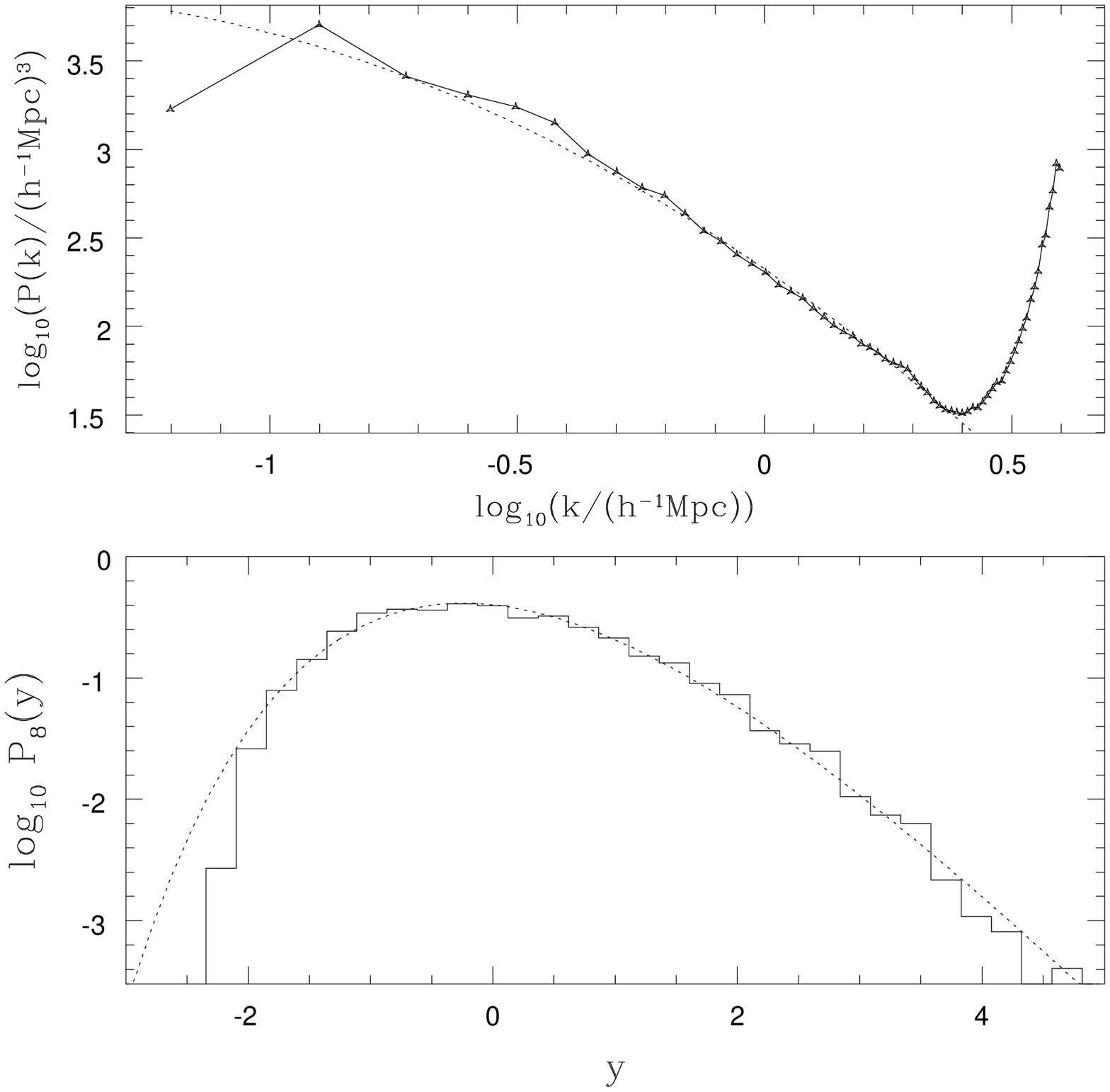,width=3.5in}}
\caption{Measured (solid lines) and fitted (dotted lines) power spectra (top
panel) and PDF (bottom panel) in the Strings-CDM model.}
\label{fig-fits-string-CDM}
\end{figure}

\begin{figure}
\centerline{\psfig{file=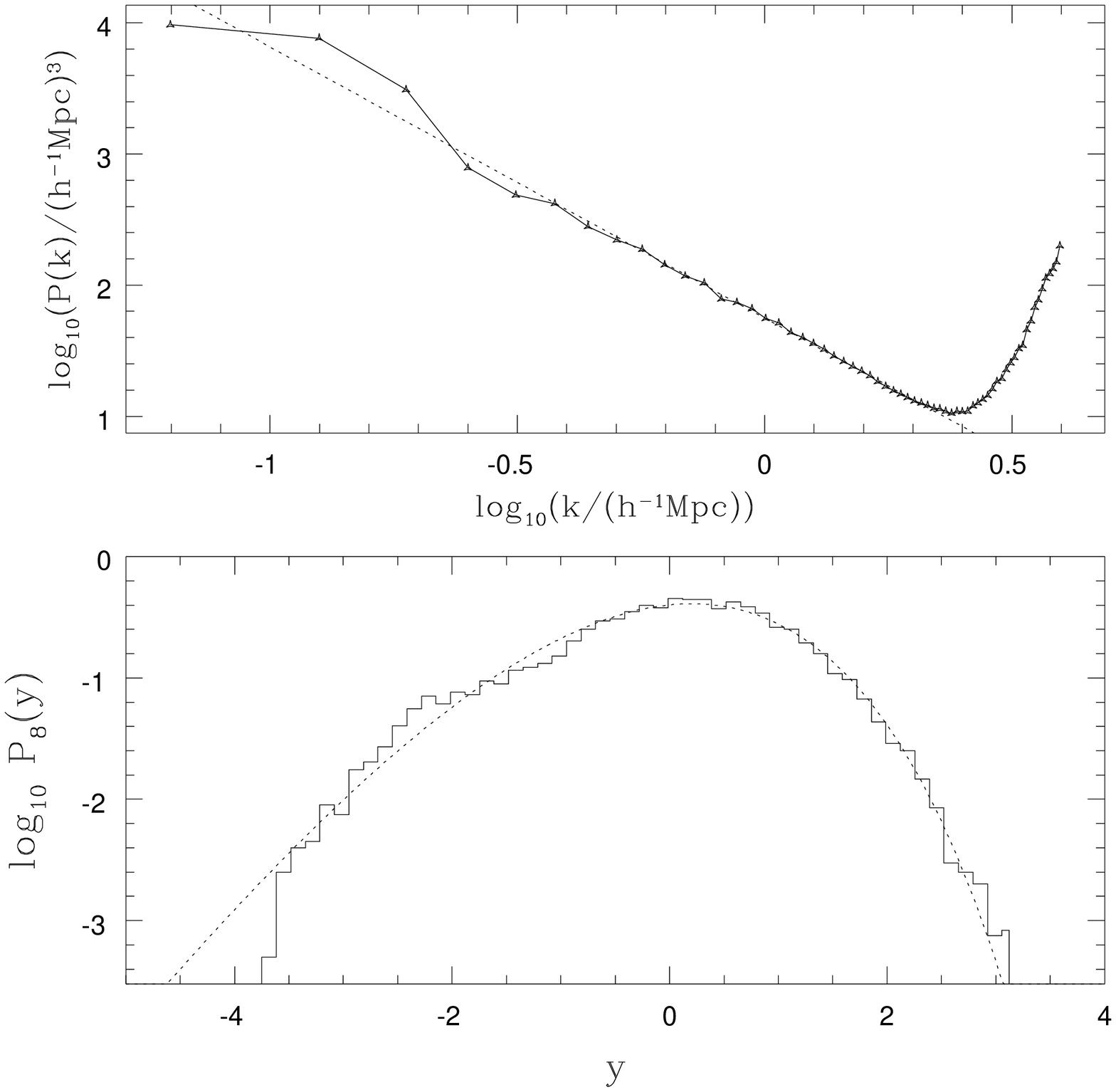,width=3.5in}}
\caption{Measured (solid lines) and fitted (dotted lines) power spectra (top
panel) and PDF (bottom panel) in the Voids model.}
\label{fig-fits-voids}
\end{figure}

The power spectrum and PDF $P_R(y)$ with $R=8h^{-1}$ Mpc (the
Lagrangian radius of a typical rich cluster) of the linear initial
conditions for each model are shown in
Figs.~\ref{fig-fits-gl-1.8},~\ref{fig-fits-icdm-2.4},~\ref{fig-fits-icdm-2.0},~\ref{fig-fits-string-HDM},~\ref{fig-fits-string-CDM},
and~\ref{fig-fits-voids}. The solid lines show the quantities
computed from the simulations, and the dotted lines show simple
analytic fits that we have used as input to the PS predictions. We fit
the power spectrum with a two parameter model:
\begin{equation}
P(k)=10^{C_1 \log_{10} k + C_2 (\log_{10} k)^2},
\end{equation}
where we compute the parameters $C_1$ and $C_2$ for each model by a
simple least squares fitting to all points $k<k_{\rm Ny}$, and $k_{\rm
Ny}=\pi L/n_{\rm p}$ is the Nyquist frequency of the box. The enhanced
power in modes with $k>k_{\rm Ny}$ is an artifact of the finite mass
resolution of the simulations; it does not affect the subsequent
evolution. For the cosmic string case, use of two parameters
significantly improves the fit to the data, but for all other models
we can obtain a good fit by setting $C_2=0$. In the ICDM-2.4 model,
the best fit power spectrum has a slope $m_\rho=-1.6$, which should be
compared with value $m_\rho=3+2m_\phi=-1.8$ which we would expect from
a theoretical calculation (see eq.~\ref{eq-mphi}). The slightly
shallower slope arises from missing power on large scales, where the
assumption of a pure power law in the scalar field $\phi$ breaks down
due to finite size effects. In the G$\Lambda$-1.8 model, the best fit
slope is $P(k)\propto k^{-1.82}$, very close to the theoretical value
of -1.8 which we have fed in.


We fit the PDF $P_R(y)$ with a log-normal distribution, translated and
normalized to have mean zero and standard deviation one, that is:
\begin{equation}
P^A_{\rm LN}(y)=\frac{C}{\sqrt{2\pi A^2}} e^{-x^2(y)/2 - |A|x(y)},
\end{equation}
where
\begin{equation}
x(y)=\frac {\ln (Cy|A|/A+B)}{|A|}
\end{equation}
with
\begin{eqnarray}
B&=&e^{A^2/2}\\
C&=&\sqrt{B^4-B^2}.
\end{eqnarray}
The log-normal distribution has one free parameter $A$, with $A>0$
giving rise to an extended tail of positive fluctuations, $A<1$ giving
rise to a suppressed tail of positive fluctuations, and Gaussianity in
the limit $A\rightarrow 0$. We fit the PDF $P_R(y)$ in each model by
finding the value of $A$ for which the number of peaks of height
3--$\sigma$ or greater is equal to that for $P_R(y)$, that is we solve
for $A$ satisfying
\begin{equation}
\int_3^\infty P_R(y)\,{\rm d}y= \int_3^\infty P^A_{\rm LN}(y)\,{\rm d}y.
\end{equation}
The log-normal PDFs fitted by this procedure also give a good fit for
all values of $y$, as demonstrated by our figures.  We find that we
can fit the scale dependence of the parameter $A$ using a quadratic
function in $R$, truncated to ensure that it does not change sign,
that is
\begin{equation}
A(R)=\left\{
\begin{array}{ll}
A_1+A_2R+A_3R^2&\ldots \frac{A_1+A_2R+A_3R^2}{(A_1+A_2R+A_3R^2)_{R=8}}>0\\
0&\ldots\mathrm{otherwise.}
\end{array}\right.
\end{equation}
For all but the Voids model, $P_R(y)$ is very nearly scale independent
over the range of scales relevant to cluster formation in our
simulation, and it is sufficient to set the parameters $A_2$ and $A_3$
to zero. In the Gaussian case, we find that the best fit value of $A$
is very slightly greater than zero, but that the predictions for
$N_{>M}$ are virtually identical whether we use this best fit value or
the choice $A=0$. Values for the parameters $\sigma_8$, $C_1$, $C_2$,
$A_1$, $A_2$, and $A_3$ used to fit the normalization, power spectrum,
and PDF in each of our models are given in Table~\ref{tab-all}.

\begin{figure}
\centerline{\psfig{file=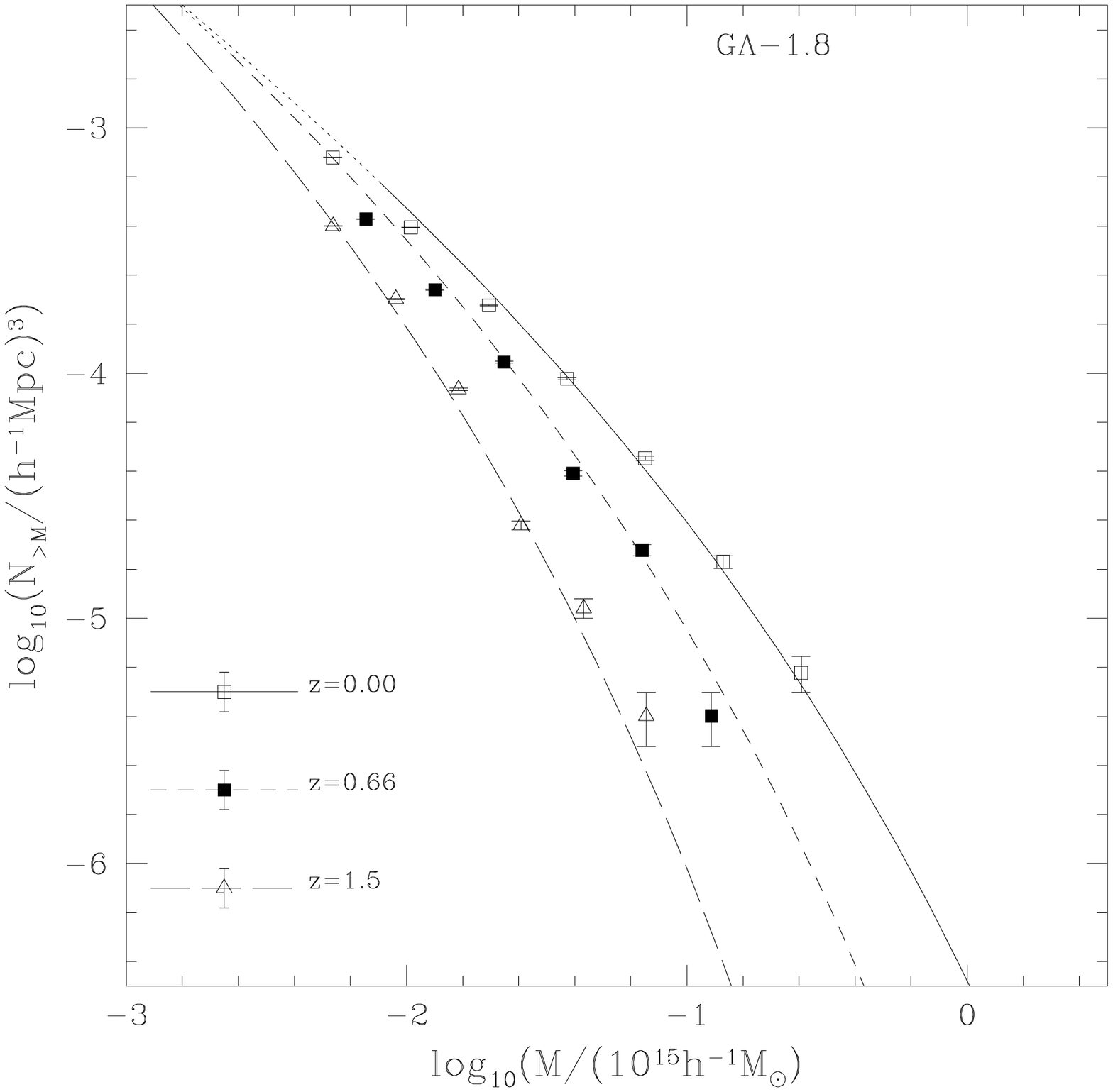,width=3.5in}}
\caption{PS prediction for the mass function (curves) and that
measured in simulations (data points) at various redshifts in the
Gaussian G$\Lambda$-1.8 model.}
\label{fig-massfunc-gl-1.8}
\end{figure}

\begin{figure}
\centerline{\psfig{file=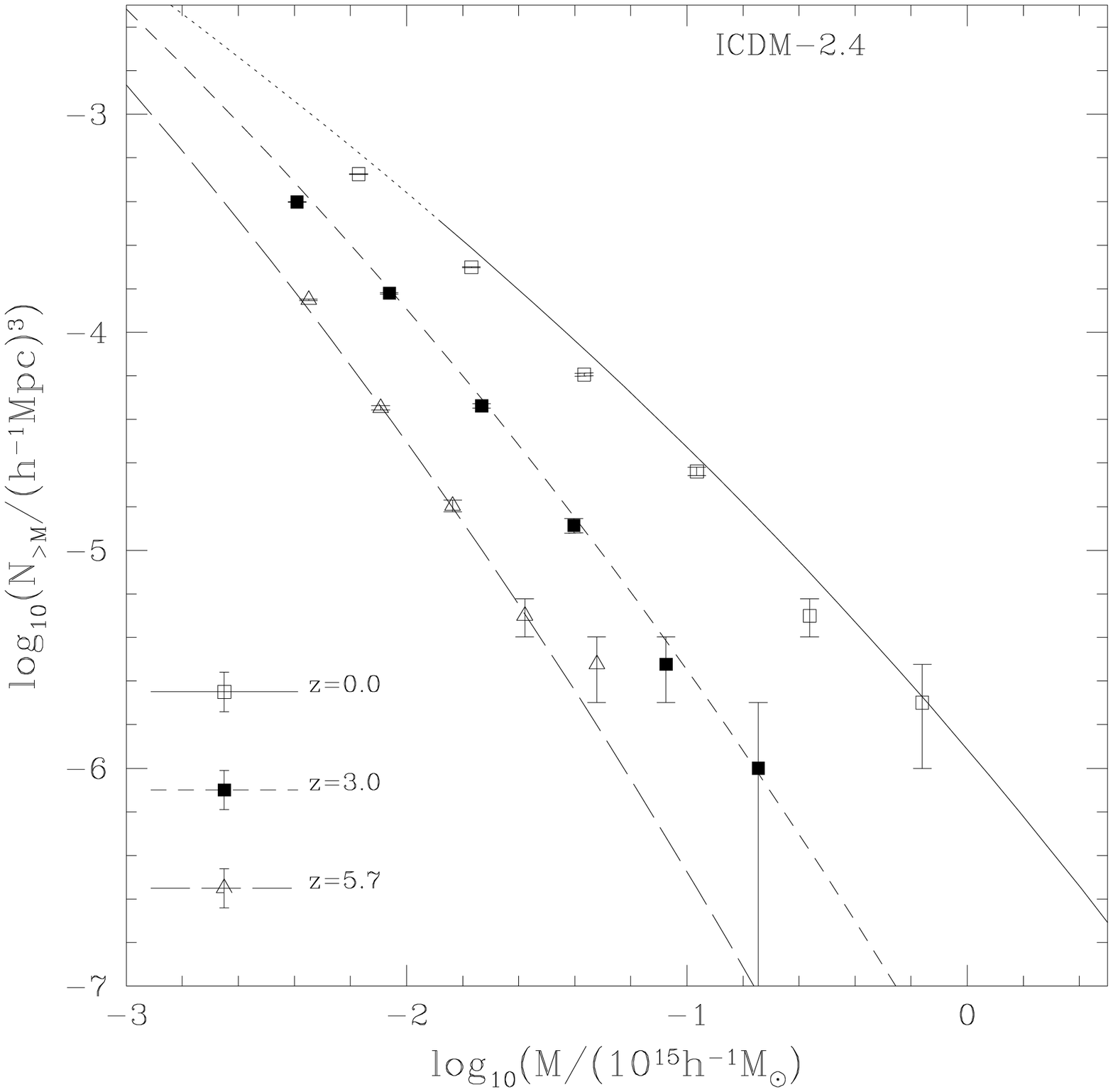,width=3.5in}}
\caption{PS prediction for the mass function (curves) and that
measured in simulations (data points) at various redshifts in the
ICDM-2.4 model.}
\label{fig-massfunc-icdm-2.4}
\end{figure}

\begin{figure}
\centerline{\psfig{file=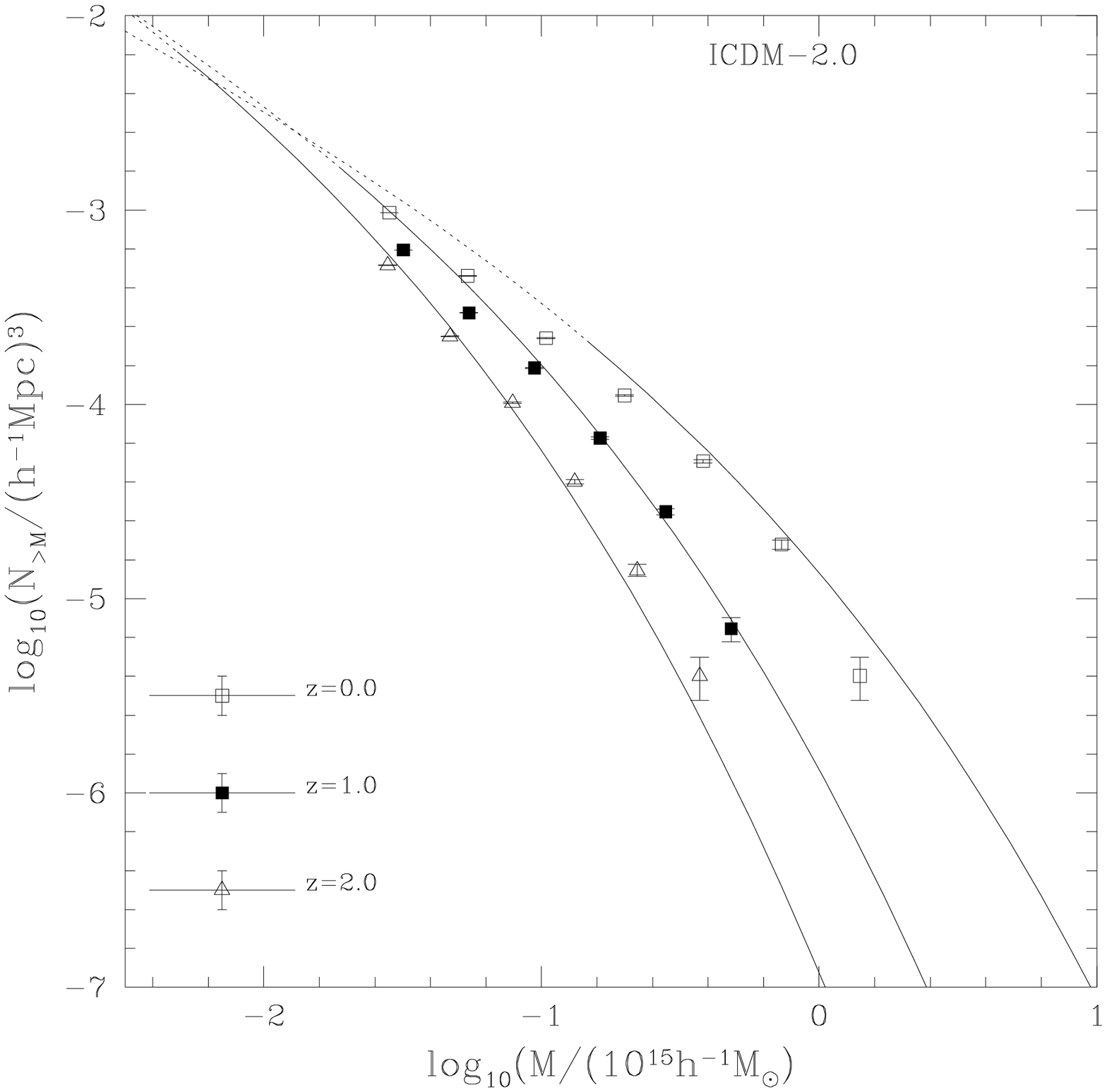,width=3.5in}}
\caption{PS prediction for the mass function (curves) and that
measured in simulations (data points) at various redshifts in the
ICDM-2.0 model.}
\label{fig-massfunc-icdm-2.0}
\end{figure}

\begin{figure}
\centerline{\psfig{file=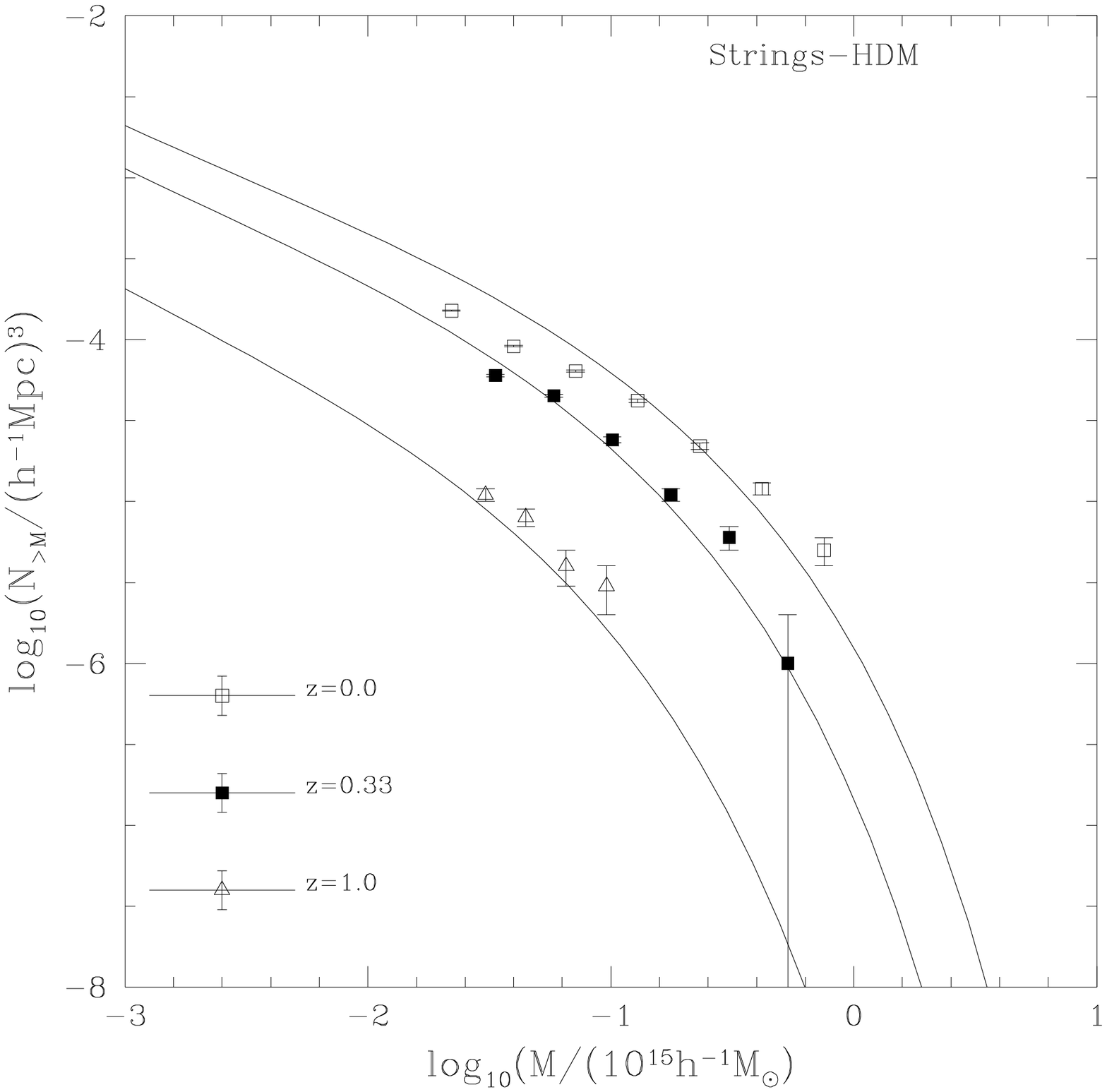,width=3.5in}}
\caption{PS prediction for mass function (curves) and that
measured in simulations (data points) at various redshifts in the
Strings-HDM model.}
\label{fig-massfunc-string-HDM}
\end{figure}

\begin{figure}
\centerline{\psfig{file=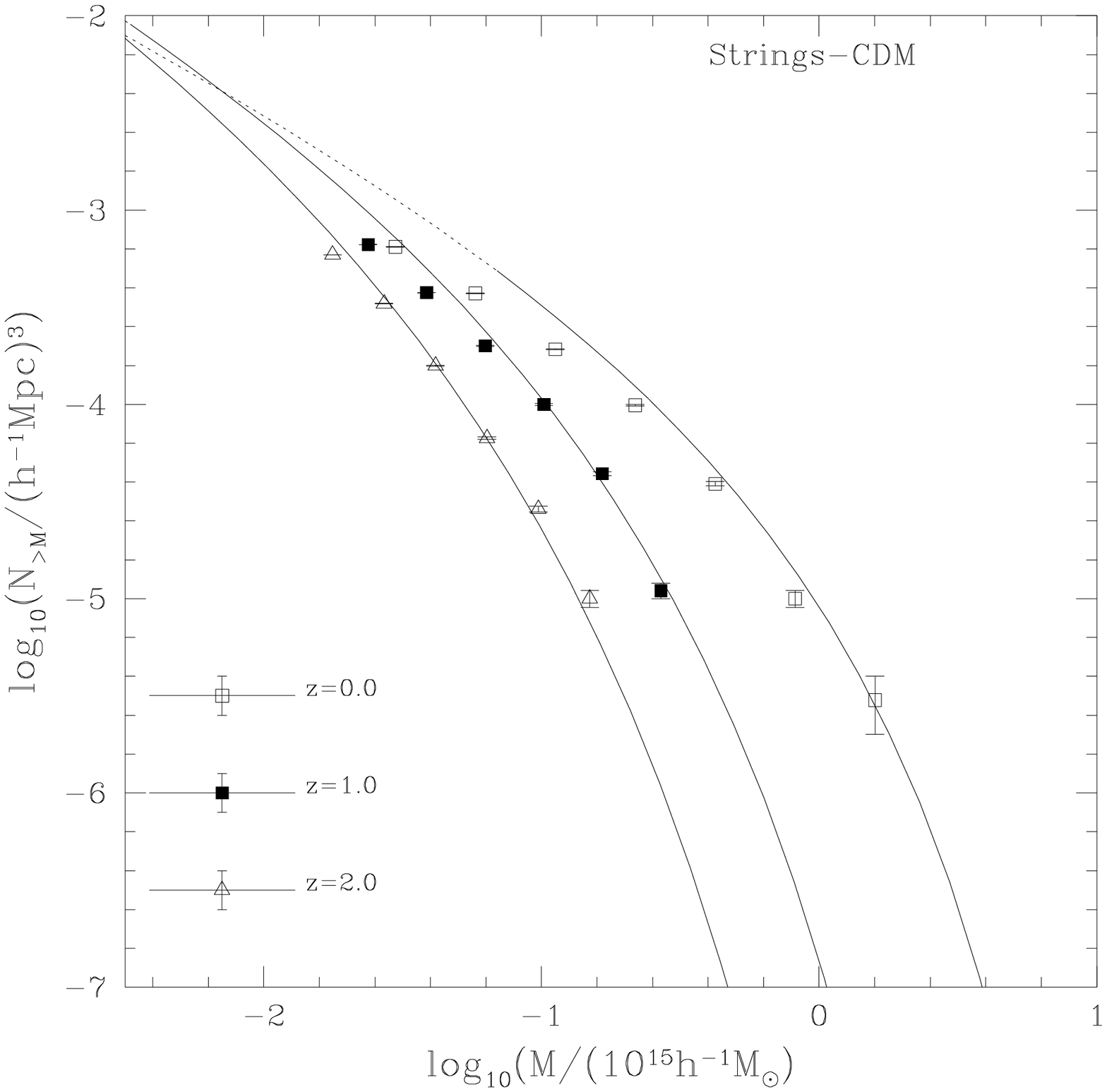,width=3.5in}}
\caption{PS prediction for mass function (curves) and that
measured in simulations (data points) at various redshifts in the
Strings-CDM model.}
\label{fig-massfunc-string-CDM}
\end{figure}

\begin{figure}
\centerline{\psfig{file=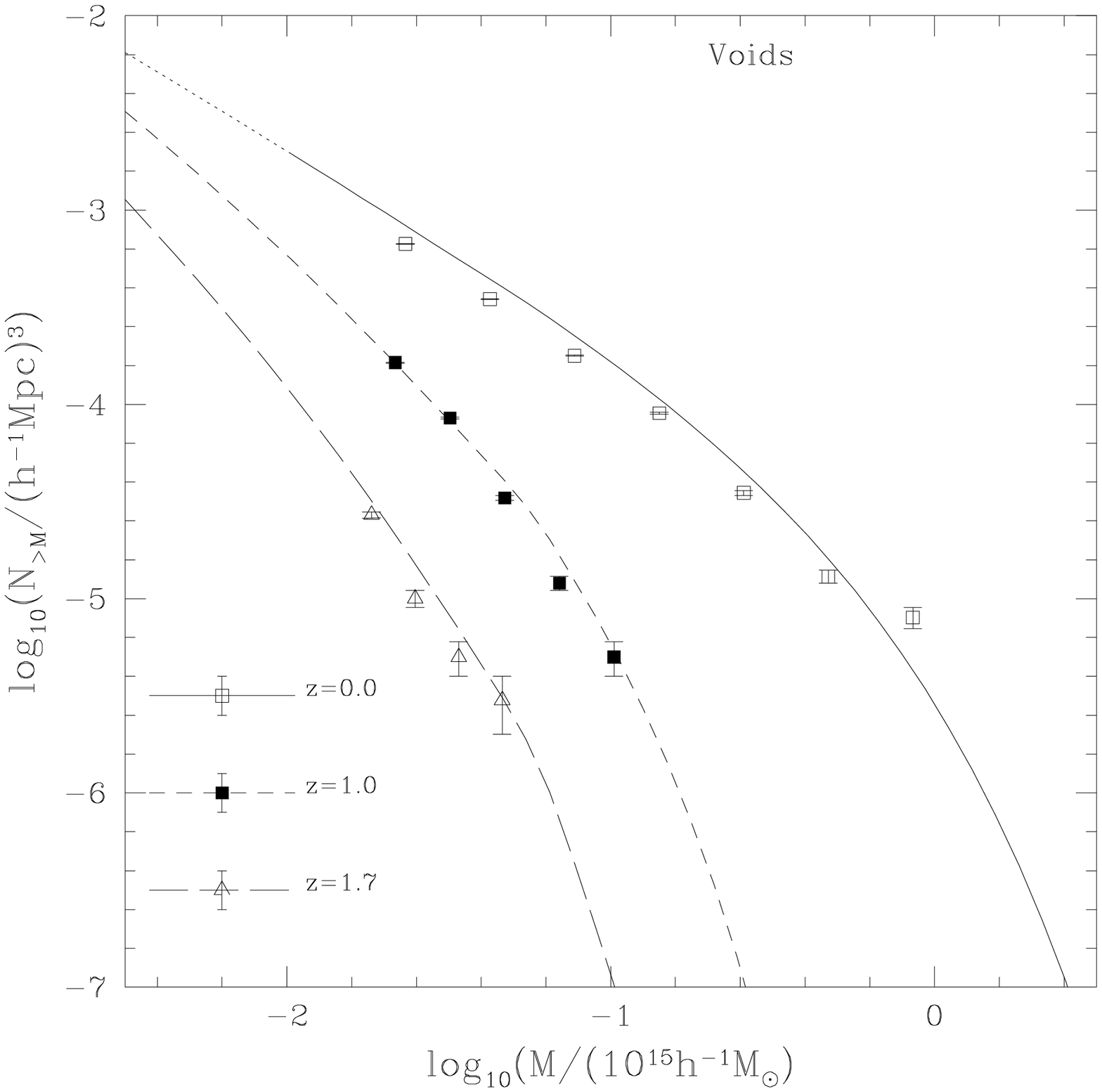,width=3.5in}}
\caption{PS prediction for the mass function (curves) and that
measured in simulations (data points) at various redshifts in the
Voids model.}
\label{fig-massfunc-voids}
\end{figure}

In
Figs.~\ref{fig-massfunc-gl-1.8},~\ref{fig-massfunc-icdm-2.4},~\ref{fig-massfunc-icdm-2.0},~\ref{fig-massfunc-string-HDM},~\ref{fig-massfunc-string-CDM},
and~\ref{fig-massfunc-voids} we compare the PS prediction $N_{>M}$ for
the mass function (curves) with the mass function $N^{S}_{>M}$
measured in the simulations (data points), for a range of output
times. For the simulations, we restrict our analysis to that portion
of the mass function $N^S_{>M}$ for which $M>16M_{\rm p}$, where $M_{\rm p}$ is
the mass of a single particle. In the Gaussian case, the PS prediction
typically gives a good fit to the mass function observed in
simulations for $M>M_{\rm NL}$, where $M_{\rm NL}$ is the typical nonlinear
mass satisfying $\delta_{\rm c}/\sigma_{R(M_{\rm NL})}=1$. We show regions for
which $M<M_{\rm NL}$ in our figures by using dotted lines to continue the
$N_{>M}$ curves.


A rough estimate of the type of errors we might expect in the PS
prediction for the non-Gaussian mass function comes from consideration
of the correction factor $f$ (see equation~\ref{eq-correction}). For
Gaussian fluctuations, the choice $f=2$ gives a good fit to the number
abundance of clusters observed in N-body simulations, and is
justified analytically by the excursion set derivation of the PS
formula (Bond et~al.\markcite{B91} 1991). For the non-Gaussian case,
we have again fixed $f$ by ensuring that the mass function accounts
for the entire mass of the universe, thus maintaining the spirit of
the PS derivation.  However, we cannot be sure {\it a priori\/}
that this this correction will still account correctly for the number
abundance of the rarest peaks. For the ICDM-2.4 model, the correction
factor corresponding to the PDF used in our calculations is $f=2.6$,
suggesting that errors of order 30\% in the predicted mass function
would not be surprising.

We quantify the goodness of fit of the predictions at each redshift by
computing the mean and the {\it rms\/} error, where the mean error
$\bar{E}$ is given by
\begin{equation}
\bar{E}=\frac{1}{j} \sum_{i=1}^{j} (N_{>M_i}-N^S_{>M_i}),
\end{equation}
and the {\it rms\/} error $\sqrt{\bar{E^{2}}}$ is given by
\begin{equation}
\sqrt{\bar{E^{2}}} = \left( \frac{1}{j} \sum_{i=1}^{j}
(N_{>M_i}-N^S_{>M_i})^2 \right)^{1/2}.
\end{equation}
Here $M_i$ is the mass corresponding to the $i^{\rm th}$ data point
and $j$ is the number of data points. Together, these two measurements
allow us to quantify both the systematic error in our predictions, as
well as the scatter about them. Errors for each redshift are quoted in
Table~\ref{tab-errors}.
\begin{table}
\centering
\begin{tabular}{cccc}
\hline 
Model& $z$ & $\bar{E}$ & $ \sqrt{\bar{E^2}}$ \\
G$\Lambda$-1.8 & 0.0 & 0.006 & 0.044  \\
               & 0.66 & 0.080  & 0.10   \\
               & 1.5 & -0.052 & 0.097   \\
ICDM-2.4       & 0.0 & 0.10   & 0.12    \\
               & 3.0 & 0.049  & 0.064   \\
               & 5.7 & -0.051 & 0.16    \\
ICDM-2.0       & 0.0 & 0.14   & 0.90    \\
               & 1.0 & 0.066  & 0.11   \\
               & 2.0 & -0.056 & 0.12    \\
Strings-HDM    & 0.0 & 0.033  & 0.18    \\
               & 0.33& -0.052 & 0.11    \\
               & 1.0 & -0.040 & 0.16    \\
Strings-CDM    & 0.0 & 0.10  & 0.12    \\
               & 1.0 & 0.081 & 0.10    \\
               & 2.0 & 0.005 & 0.065    \\
Voids          & 0.0 & 0.047  & 0.13    \\
               & 1.0 & 0.057  & 0.084   \\
               & 1.7 & 0.11   & 0.13    \\
\hline
\end{tabular}
\vspace{10pt}
\caption{Mean and {\it rms\/} errors in the predicted mass functions at
each redshift.}
\label{tab-errors}
\end{table}
In computing these errors we do not attempt to
fit for points for which $M<M_{\rm NL}$, since the PS formalism is known to
break down for these masses in the Gaussian case. We see that typical
errors are less than of order $0.1$ in
$\log_{10}{(N_{>M}/(h^{-1}{\rm Mpc})^3)}$. The worst systematic error is for the
Voids model at $z=1.7$, where $\bar{E}=0.11$. The 
worst {\it rms\/} error is for the Strings-HDM model at $z=0$, where 
$\sqrt{\bar{E^{2}}}=0.18$. Even in these worst cases, the errors are of
order or smaller than typical measurement errors in $N_{>M}$, and many
orders of magnitude smaller than the amount by which predictions for
different models can differ.
  
\section{Conclusions}
\label{sec-conclusions} 
We have carried out N-body simulations of a number of models with
non-Gaussian initial conditions, and verified that a modified version
of the Press-Schechter formalism is able to give a good fit to the
observed evolution of the cluster number abundance over a wide range
of redshifts. While we have only tested the fit for a finite number
of models, the models span a range of possible types of non-Gaussian
behaviour. The ICDM and string models have a skew positive PDF, and
typically give rise to more clusters of a given mass than would be
expected in the corresponding Gaussian case. The Voids model has a
skew negative PDF, and typically gives rise to fewer clusters than
would be expected in the corresponding Gaussian case. The ICDM model
is generated purely by a local transformation on a Gaussian field, so
might be thought most likely to agree with the PS prediction. The
Voids model, on the other hand, is generated by a more complex non-local
transformation of a Gaussian random field, and the strings models are
generated by an entirely different process altogether.

Quantitatively, the fits to the predicted mass function are slightly
worse in the non-Gaussian models than in the Gaussian one. Typical
fits in the non-Gaussian models have {\it rms\/} errors in
$\log_{10}(N_{>M}/(h^{-1}{\rm Mpc})^3)$ of $\sqrt{\bar{E^2}}\simeq 0.1$,
corresponding to an error of roughly 25\% in $N_{>M}$. The worst fit
is in the $z=0$ output of a strings model, where the {\it rms\/} error
is $\sqrt{\bar{E^2}}=0.18$, or roughly 50\% in $N_{>M}$. These errors
are of order or smaller than typical observational uncertainties in
the determination of the mass function, and considerably smaller than
the amount by which predictions for different models may differ. 

Given uncertainties at this level, our results make it possible to use
observations of cluster evolution to place strong constraints on
non-Gaussianity in the primordial universe. The constraints from
current cluster data are outlined in detail in Robinson
et~al.\markcite{Retal} (in preparation). Given an independent measure
of the matter density of the universe $\Omega_{\rm m}$ (which we can
hope to gain in the near future from the combination of supernovae and
CMB observations), cluster evolution directly constrains the
probability distribution function (PDF) of primordial fluctuations in
the universe. If we assume $\Omega_{\rm m}=1$, Robinson
et~al.\markcite{Retal} show that current data detects non-Gaussianity,
and constrains the primordial PDF to have at least 2.5 times as may
3--$\sigma$ peaks as a Gaussian distribution. If we assume
$\Omega_{\rm m}=0.3$, current data is consistent with Gaussian
fluctuations, with the PDF constrained to have no more than 9 times as
may 3--$\sigma$ peaks as a Gaussian distribution. Taken together with
the substantial increase of data on high redshift clusters we can
expect in the near future, the techniques described here can provide
powerful, model independent constraints on non-Gaussianity in the
primordial universe.

\section{acknowledgements}
We would like to thank Marc Davis, Eric Gawiser, Joseph Silk, and
Jochen Weller for helpful and stimulating discussions. We would also
like to thank Pedro Ferreira for the use of his string evolution
code. This work has been supported in part by grants from the NSF,
including grant AST95-28340.


\begin{thebibliography}{}

\def \prl {PRL}
\def \apj {ApJ}
\def \aap  {A\&A}
\def \mnras {MNRAS}
\def \prb {PRB}
\def \prd {PRD}

\bibitem{ABR} 
Albrecht A., Battye R.~A., Robinson, J., 1997, \prl, 79, 4736

\bibitem{Aetal97}
Allen B., Caldwell R.~R., Dodelson S., Knox L., Shellard
E.~P.~S., Stebbins A., 1997, \prl, 79, 2624

\bibitem{AB94}
Amendola A., Borgani S., 1994, \mnras, 266, 191

\bibitem{A091}
Amendola A., Occhionero F., 1993, \apj, 413, 39

\bibitem{ASWA}
Avelino P.~P., Shellard E.~P.~S., Wu J.~H.~P., Allen, B., 1998, \prl,
81, 2008




\bibitem{BRA}
Battye R.~A., Robinson J., Albrecht A., 1998, \prl, 80, 4847


\bibitem{B91}
Bond J.~R., Cole S., Efstathiou G., Kaiser N.. 1991, \apj, 379,
440

\bibitem{BCMP94}
Borgani S., Coles P., Moscardini L., Maniolis P., 1994, \mnras, 266, 524

\bibitem{BSO95}
Brieu P.~P., Summers F.~J., Ostriker J.~P.,  1995, \apj, 453, 566

\bibitem{COS} Chiu W.~A., Ostriker J.~P., Strauss M.~A., 1997,
\apj, 494, 479







\bibitem{F95}
Ferreira P., 1995, PhD Thesis, Imperial College





\bibitem{K76}
Kibble T.~W.~B., 1976, J.~Phys., A9, 1387

\bibitem{L91}
La D., 1991, \prb, 265, 232

\bibitem{LC}
Lacey C., Cole S., 1993, \mnras, 262 627



\bibitem{MJW}
Mo H.~J., Jing Y.~P., White S.~D.~M., 1996, \mnras, 
284, 189



\bibitem{PST}
Park C., Spergel D.~N., Turok N., 1991, \apj, 372, L53

\bibitem{P83}
Peebles P.~J.~E., 1983, \apj, 274, 1

\bibitem{P97}
Peebles P.~J.~E., 1997, \apj, 483, L1

\bibitem{P98a}
Peebles P.~J.~E., 1998, astro-ph/9805194

\bibitem{P98b}
Peebles P.~J.~E., 1998, astro-ph/9805212

\bibitem{PS}
Press W.~H., Schechter P., 1974, \apj, 187, 425

\bibitem{RGS}
Robinson J., Gawiser E., Silk J., 1998, astroph/9805181 


\bibitem{SmV}
Smith A.~G.,  Vilenkin A., 1987, \prd, 36, 990


\bibitem{vdB}
van de Bruck C., 1998, astroph/9810409

\bibitem{VandS}
Veeraraghavan S., Stebbins A., \apj, 365, 37


\bibitem{VS}
Vilenkin A., Shellard E.~P.~S., 1994, 
Cosmic strings and other topological defects. Cambridge
Univ. Press, Cambridge


\bibitem{WC}
Weinberg D.~H., Cole S., 1992, \mnras, 259, 652

\bibitem{Z70}
Zeldovich, Y.  1970, \aap, 5, 84

\end{thebibliography}
\end{document}